\documentstyle[11pt]{article}

\def\kon#1#2{\vbox{\halign{##&&##\cr\lower4pt
\hbox{$\scriptscriptstyle\vert$}\hrulefill &\hrulefill\lower4pt
\hbox{$\scriptscriptstyle\vert$}\cr $#1$&$#2$\cr}}}

\def\fii{\varphi}
\def\al{\alpha}
\def\be{\beta}

\def\d{\partial}
\def\=d{\,{\buildrel\rm def\over =}\,}

\def\sqr#1#2{{\vcenter{\vbox{\hrule height.#2pt\hbox{\vrule width.
#2pt height#1pt \kern#1pt \vrule width.#2pt}\hrule height.#2pt}}}}
 
\def\eps{\varepsilon}

\def\la{\lambda}

\def\te{\vartheta}
\def\fii{\varphi}

\begin{document}

\title{Exact gravitational lensing and rotation curve}
\author{\\G\"unter Scharf and Gerhard Br\"aunlich 
\footnote{e-mail: scharf@physik.unizh.ch}
\\ Institut f\"ur Theoretische Physik, 
\\ Universit\"at Z\"urich, 
\\ Winterthurerstr. 190 , CH-8057 Z\"urich, Switzerland}

\date{}

\maketitle\vskip 3cm

\begin{abstract} Based on the geodesic equation in a static spherically symmetric
metric we discuss the rotation curve and gravitational lensing. The rotation curve
determines one function in the metric without assuming Einstein's equations. Then
lensing is considered in the weak field approximation of general relativity. 
From the null geodesics we derive the lensing equation and corrections to it.
\end{abstract}

\newpage

\section{Introduction}

As long as the dark matter problem is open there is a non-zero probability that
general relativity might not hold on the scale of galaxies [1-4]. Therefore a direct
test on this scale is highly desired. It is the purpose of this paper to show how
such a test is possible, if kinematical and lensing data of the galaxy are available.
The idea is the following: the rotation curve determines part of the metric {\it without}
assuming Einstein's field equations, only the geodesic equation is used. Then lensing
can be calculated on the basis of the weak field approximation to general relativity and
checked for consistency. In contrast to the usual way of analyzing the data no model
for the galaxy must be constructed. This offers the possibility to test the basic physics.

We treat lensing by means of the geodesic equation as well.
By computing the null geodesics we derive the lensing equation and we find corrections to it.
Even if these corrections were not needed for the analysis of present day lensing data,
they have to be under control for all eventualities.
We only consider the static spherically symmetric case here in order to make the argument as
simple as possible.

\section{Geodesic flow and rotation curve}

We consider a static spherically symmetric metric which we write in the form
$$ds^2=e^\nu c^2dt^2-e^\la dr^2-r^2(d\te^2+\sin^2\te d\fii^2)\eqno(2.1)$$
where $\nu$ and $\la$ are functions of $r$ only. We take the coordinates
$x^0=ct$, $x^1=r$, $x^2=\te$, $x^3=\fii$ such that
$$g_{00}=e^\nu,\quad g_{11}=-e^\la$$
$$g_{22}=-r^2,\quad g_{33}=-r^2\sin^2\te\eqno(2.2)$$
and zero otherwise. The components with upper indices are the inverse of
this. The determinant comes out to be
$$g={\rm det}g_{\mu\nu}=-e^{\nu+\la}r^4\sin^2\te.\eqno(2.3)$$

Let us recall the Christoffel symbols for the metric (2.1)
from the appendix of [3]
$$\Gamma^0_{10}={1\over 2}\nu',\quad \Gamma^1_{00}={1\over 2}\nu'e^{\nu-\la},
\quad \Gamma^1_{11}={1\over 2}\la',\quad \Gamma^1_{22}=-re^{-\la}\eqno(2.4)$$
$$\Gamma^1_{33}=-re^{-\la}\sin^2\te,\quad \Gamma^2_{12}={1\over r},
\quad \Gamma^2_{33}=-\sin\te\cos\te,\quad \Gamma^3_{13}={1\over r},
\quad \Gamma^3_{23}=\cot\te$$
and zero otherwise, the prime denotes the derivative with respect to $r$
always. The geodesic equation is given by
$${d^2x^\al\over ds^2}+\Gamma^\al_{\be\gamma}{dx^\be\over ds}
{dx^\gamma\over ds}=0.\eqno(2.5)$$
The origin of our reference frame is in the center of the galaxy.
We consider geodesics in the plane $\theta=\pi/2$, then we must solve the
following three equations
$${d^2ct\over ds^2}+\nu'{d\, ct\over ds}{dr\over ds}=0\eqno(2.6)$$
$${d^2r\over ds^2}+{\nu'\over 2}e^{\nu-\la}\Bigl({d\, ct\over ds}\Bigl)^2+
{\la'\over 2}\Bigl({dr\over ds}\Bigl)^2-re^{-\la}\Bigl({d\fii\over ds}
\Bigl)^2=0\eqno(2.7)$$
$${d^2\fii\over ds^2}+{2\over r}{dr\over ds}{d\fii\over ds}=0.\eqno(2.8)$$

Multiplying (2.6) by $\exp\nu$ we find
$${\d\over ds}\Bigl(e^\nu{d\,ct\over ds}\Bigl)=0$$
so that
$$e^\nu{d\,ct\over ds}={\rm const.}= a$$
$${d\,ct\over ds}=ae^{-\nu}.\eqno(2.9)$$
Next multiplying (2.8) by $r^2$ we get
$$r^2{d\fii\over ds}={\rm const.}=J$$
where $J$ is essentially the conserved angular momentum, hence
$${d\fii\over ds}={J\over r^2}.\eqno(2.10)$$

Finally, substituting (2.9) and (2.10) into (2.7) and multiplying by
$2(\exp\la)$ $\times dr/ds$ we obtain
$${d\over ds}\Bigl[e^\la\Bigl({dr\over ds}\Bigl)^2-a^2e^{-\nu}+
{J^2\over r^2}\Bigl]=0.\eqno(2.11)$$
Consequently, the square bracket is equal to another constant $=b$.
Then the resulting differential equation can be written as
$$\Bigl({dr\over ds}\Bigl)^2=a^2e^{-(\la+\nu)}+e^{-\la}\Bigl(b-
{J^2\over r^2}\Bigl).\eqno(2.12)$$

To obtain the connection with the rotation curve which is an important
astronomical observable, we remember the definition of the unitary
4-velocity
$$u^\al={dx^\al\over ds}.$$
The term unitary indicates that $u^\al$ has invariant length 1:
$$u^2=g_{\al\be}{dx^\al\over ds}{dx^\be\over ds}={(ds)^2\over (ds)^2}
=1.\eqno(2.13)$$
In our case $u^\al$ is equal to
$$u^\al=\Bigl({d\,ct\over ds}, {dr\over ds},0,{d\fii\over ds}\Bigl).
\eqno(2.14)$$
Using (2.9) (2.10) and (2.12) we easily see that
$$u^2=-b=1\eqno(2.15)$$
which by (2.13) fixes the constant of integration $b=-1$.

Clearly the last constant of integration $a^2$ must be related to the
geometry of the geodesics. To see this we consider the
streamlines $r=r(\fii)$. Deviding (2.12) by $J=r^2d\fii/ds$ we obtain
$$\Bigl({1\over r^2}{dr\over d\fii}\Bigl)^2={a^2\over J^2}e^{-(\la+\nu)}
+e^{-\la}\Bigl({b\over J^2}-{1\over r^2}\Bigl).\eqno(2.16)$$
Introducing the variable
$$w(\fii)={1\over r(\fii)},\eqno(2.17)$$
we write the equation in the form
$$\Bigl({dw\over d\fii}\Bigl)^2={a^2\over J^2}e^{-(\la+\nu)}
+e^{-\la}\Bigl({b\over J^2}-w^2\Bigl).\eqno(2.18)$$
To compare this equation with Newtonian dynamics we use the expansion
of the metric for large $r$: 
$$e^{-(\la+\nu)}=1+O(r^{-2}),\quad e^{-\la}=1-{r_s\over r}+O(r^{-2}).$$
Here
$$r_s={2GM\over c^2}\eqno(2.19)$$
is theSchwarzschild radius in case of a point mass. Then to order $1/r$ we have
$$\Bigl({dw\over d\fii}\Bigl)^2+w^2={a^2+b\over J^2}-
r_Sw\Bigl({b\over J^2}-w^2\Bigl).\eqno(2.20)$$
In Newtonian mechanics the bounded streamlines are ellipses
$$\tilde w={1\over r}={1\over p}(1+e\cos\fii),\eqno(2.21)$$
where $p$ and $e$ are parameter and eccentricity of the ellipse.
$p$ is connected with the non-relativistic angular momentum $\tilde J$ by
$${\tilde J^2\over p}=GM.\eqno(2.22)$$
The Newtonian equation which corresponds to (2.20) now reads
$$\Bigl({d\tilde w\over d\fii}\Bigl)^2+\tilde w^2={e^2-1\over p^2}+
{2\over p}\tilde w.\eqno(2.23)$$
Comparing the coefficients in (2.20) and(2.23) we first find
$$-r_S{b\over J^2}={2\over p}.$$
By (2.22) and $\tilde J=cJ$ this gives $b=-1$ in agreement with (2.15). 
Secondly, from
$${a^2+b\over J^2}={e^2-1\over p^2}=-{1\over p\tilde a},$$
where $\tilde a$ is the big half-axis of the ellipse, we obtain
by (2.22)
$$a^2=1-{GM\over c^2\tilde a}=1-{r_S\over 2\tilde a}.\eqno(2.24)$$
This shows that $a^2$ is connected with the big half-axis of the
Kepler ellipse.

The 3-velocity $\vec v$ which is measured by astronomers is defined as
$$\vec v=\Bigl({dx^1\over dt}, {d x^2\over dt},{d x^3\over dt}\Bigl).
\eqno(2.25)$$
Using
$${ds\over dt}={c\over a}e^\nu\eqno(2.26)$$
we can calculate
$$\vec v^2=\vec u^2\Bigl({ds\over dt}\Bigl)^2,\eqno(2.27)$$
where $\vec u^2$ is the spatial part in (2.13). Since our metric is
diagonal it is simply given by
$$-\vec u^2=\sum_{j=1}^3g_{jj}{dx^j\over ds}{dx^j\over ds}=1-a^2e^{-\nu}.\eqno(2.28)$$
By (2.26) we now get the desired velocity squared
$$\vec v^2=c^2\Bigl(e^\nu-{e^{2\nu}\over a^2}\Bigl).
\eqno(2.29)$$

As a check we determine the asymptotic behavior for large $r\gg r_S$.
Assuming circular motion ($\tilde a=r$) and using $1/a^2=1+GM/c^2r$ we find
$$\vec v^2\to c^2\Bigl(-\nu (r)-{GM\over c^2r}+ O(r^{-2})\Bigl).\eqno(2.30)$$
Since $\nu=-r_s/r$
we arrive at
$$\vec v^2\to {GM\over r},\eqno(2.31)$$
$M$ is the total mass (normal plus dark). This agrees with Newtonian 
dynamics (Kepler's third law).
Summing up, the relation between observational quantities and theory is very
direct. The rotation curve $v(r)$ gives the metric function $\nu(r)$ by
solving the quadratic equation (2.29)
$$e^\nu={a^2\over 2}\Bigl(1\pm\sqrt{1-{4\over a^2}{v^2\over c^2}}
\Bigl).\eqno(2.32)$$
For velocities $v\ll c$ and $r\gg r_s$ this simplifies to
$$e^\nu={1\over 2}\pm {1\over 2}\sqrt{1-4{v^2\over c^2}}.\eqno(2.33)$$

\section{Null geodesics and lensing}

In the case of null geodesics describing light rays the integration constant $b$ 
in the geodesic equation (2.16) must be 0
$${1\over r^4}\Bigl({dr\over d\fii}\Bigl)^2=e^{-\la}\Bigl({a^2\over J^2}e^{-\nu}-
{1\over r^2}\Bigl).\eqno(3.1)$$
In the lensing problem one uses the weak field approximation to general relativity
$$e^{-\nu}=1-{2U(r)\over c^2},\quad e^{-\la}=1+{2U(r)\over c^2}\approx e^\nu,
\eqno(3.2)$$
where $U(r)$ is the gravitational potential. The latter can be obtained from the
rotation velocity according to (2.32) or (2.33). Expanding the square root in (2.33)
for $v^2\ll c^2$ we get the very simple result
$$U(r)=-{v^2\over c^2},\eqno(3.3)$$
we see that the plus sign in (2.33) must be used. Introducing the quantity
$$d={J\over a}\eqno(3.4)$$
in (3.1), the following first order equation remains to be solved
$$\Bigl({dr\over d\fii}\Bigl)^2={r^4\over d^2}\Bigl(1-{4U^2\over c^4}\Bigl)-
\Bigl(1+{2U\over c^2}\Bigl)r^2.\eqno(3.5)$$
The meaning of $d$ becomes clear when we consider the trivial solution for $U=0$:
$$r={d\over\sin\fii}.$$
It describes a straight line with distance $d$ from the origin in polar coordinates
(fig.1). After inversion the equation (3.5) can simply be solved by quadrature:
$${d\fii\over dr}={\pm 1\over r\sqrt{{r^2\over d^2}(1-{4U^2\over c^4})-1-{2U\over c^2}}}
\eqno(3.6)$$
The sign herein depends on the branch of the geodesic to be calculated.

In case of a point-mass (Schwarzschild) lens we have
$$U(r)=-{GM\over r}\eqno(3.7)$$
and from (3.6) we get an elliptic integral for the polar angle $\fii(r)$:
$$\fii(r)-\fii_0=\pm d\int\limits_{r_0}^r{dr\over\sqrt{r^4-r^2(d^2+r_s^2)+rr_sd^2}},
\eqno(3.8)$$
where we have again used the Schwarzschild radius $r_s$ (2.19). To reduce this integral to
Legendre's normal form we need the four zeros $a_1, a_2, a_3, a_4$ of the quartic under
the square root. We have the following four real roots
$$a_1=d\Bigl(\sqrt{1+{\eps^2\over 4}}-{\eps\over 2}\Bigl),\quad a_2=\eps d
,\eqno(3.9)$$
$$a_3=0,\quad a_4=d\Bigl(-\sqrt{1+{\eps^2\over 4}}-{\eps\over 2}\Bigl),\eqno(3.10)$$
where
$$\eps={r_s\over d}\eqno(3.11)$$
is a small parameter. It is convenient to expand everything in powers of $\eps$:
$$a_1=d(1-{\eps\over 2}+{\eps^2\over 8}),\quad a_2=\eps d,\quad a_3=0,\quad
a_4=-d(1+{\eps\over 2}+{\eps^2\over 8})\eqno(3.12)$$
up to $O(\eps^3)$. The integral (3.8) is an incomplete elliptic integral of the first kind
$F(\Phi,k)$ where the parameter $k$ is given by
$$k^2={(a_2-a_3)(a_1-a_4)\over (a_1-a_3)(a_2-a_4)}=2\eps(1-\eps)\eqno(3.13)$$
(see [4], vol.II, p.310).

As  first application we compute the Einstein deflection angle and the correction to it.
The origin of our coordinate system is at the mass $M$, polar axis goes from $M$ to the
observer (see fig.). We integrate (3.8) from the apex $r_0$ to infinity which gives us
the deflection angle $\fii_\infty-\pi/2$. The apex is defined by the condition
$${dr\over d\fii}=0$$
which gives $r_0=a_1$. Then we obtain
$$\fii_\infty-{\pi\over 2}=\mu dF(\Phi_\infty,k),\eqno(3.14)$$
where the Jacobian $\mu$ is equal to
$$\mu={2\over\sqrt{(a_3-a_1)(a_4-a_2)}}={2\over d}\Bigl(1-{\eps\over 2}+{\eps^2\over 2}
\Bigl)\eqno(3.15)$$
The argument $\Phi_\infty$ follows from
$$\sin^2\Phi_\infty={a_4-a_2\over a_4-a_1}={1\over 2}\Bigl(1+{3\over 2}\eps-{1\over 8}
\eps^2\Bigl)={1\over 2}(1-\cos (2\Phi_\infty))\eqno(3.16)$$ 
(see [4], vol.II, p.310). The elliptic integral can be expanded for small $k$ as follows
$$F(\Phi,k)=\Phi+{k^2\over 4}(\Phi-{1\over 2}\sin 2\Phi)+O(k^4)\eqno(3.17)$$
(see [4], vol.II, p.313).
This finally gives
$$\fii_\infty=-\eps+\eps^2\Bigl({\pi\over 8}-{7\over 8}\Bigl).\eqno(3.18)$$
The $\eps=r_s/d$ is Einstein's result.

Next we want to derive the lens equation. In this problem the observer is not at infinity
but in a finite distance $D_d$ from the lens. The light source is at a distance $D_{ds}$
at the other side of the lens and an amount $\eta$ off the optical axis (see fig.); we use
the same notation as in [5]. We have now to compute the null geodesics from the source at
distance $D_{ds}$ through the apex $r=r_0,\te=\pi/2$ to the observer at distance $D_d$.
Then the polar angle $\tilde\beta={\eta\over D_{ds}}$ of the source follows from
$$\pi-\tilde\beta=\mu d[F(\Phi_{ds},k)+F(\Phi_d,k)].\eqno(3.19)$$
Here the angle $\Phi_d$ is given by
$$\sin^2\Phi_d={a_4-a_2\over a_4-a_1}{D_d-a_1\over D_d-a_2}=$$
$$={1\over 2}(1-\cos(2\Phi_d))={1\over 2}\Bigl[1+{3\over 2}\eps-\Theta(1-{17\over 8}\eps^2) 
+\Theta^2\eps\Bigl]\eqno(3.20)$$
where $\Theta=d/D_d$ is the angle under which the observer sees the source. Then from (3.17)
we obtain
$$F(\Phi_d,k)={\pi\over 4}+\eps\Bigl({1\over 2}+{\pi\over 8}\Bigl)-{\Theta\over 2}
-{\eps\Theta\over 4}+\eps^2\Bigl({5\over 8}-{\pi\over 8}\Bigl).\eqno(3.21)$$
$F(\Phi_{ds},k)$ is given by the same formula with $\Theta$ substituted by $\al=d/D_{ds}$.
Now we find from (3.19)
$$-\tilde\beta=-\Theta-\al+2\eps+\eps^2\Bigl({3\over 3}-{\pi\over 4}\Bigl)+O(\eps^3)
.\eqno(3.22)$$
The lens equation is usually written in terms of the angles
$$\beta=\tilde\beta{D_{ds}\over D_s},\quad\Theta={r_s\over D_d\eps},\quad
\al=\Theta{D_d\over D_{ds}}.\eqno(3.23)$$
Then (3.22) gives the following lens equation with corrections 
$$\beta=\Theta-2{D_{ds}\over D_s}{r_s\over D_d\Theta}
-\Bigl({3\over 2}-{\pi\over 4}\Bigl){D_{ds}\over D_s}
\Bigl({r_s\over D_d\Theta}\Bigl)^2\eqno(3.24)$$
for the point-mass lens. The first three terms are the leading order standard result
([6], p.27). The third term can be written in terms of the deflection angle (3.18) as it
is usually done. Comparing the corrections in (3.24) with those in (3.18) we find no
direct correspondence. That means the lens equation as usually written by means of the
scaled deflection angle ([5]. p.21) is the leading approximation only.

Now we turn to the formulation of the lens equation in an arbitrary spherically symmetric metric.
From (3.6) we have the following integral for the polar angle of the null geodesics:
$$\fii(r)-{\pi\over 2}=\pm d\int\limits_{r_0}^r{dr\over r\sqrt{r^2(1-u^2)-(1+u)d^2}},
\eqno(3.25)$$
where we have introduced the dimensionless gravitational potential
$$u(r)={2\over c^2}U(r).\eqno(3.26)$$
Here $r_0$ is the apex and it is important to note that only the potential values
for $r\ge r_0$ contribute. For $u=0$ the trivial lens equation $\beta=\Theta$ comes
out, this follows from (3.24) for $r_s=0$. For $u\ne 0$ but $\vert u\vert\ll 1$ the
modification of the result comes from the neighborhood of the apex. It is therefore
good enough to expand the potential $u(r)$ in the vicinity of $r=r_0$. For this purpose
we use the beginning of the multipole expansion
$$u(r)=c_0+{c_1\over r}+{c_2\over r^2}.\eqno(3.27)$$
The constant term is necessary in view of the flat rotation curves; note that the
potential has an absolute normalization in (3.2). It is unimportant that (3.27) breaks
down for small $r$ because we need $r\ge r_0$ only. 

With the three terms in (3.27) we get an elliptic integral of the first kind again:
$$\fii(r)-{\pi\over 2}=\pm d\int\limits_{r_0}^r{dr\over \sqrt{G(r)}}=\pm 
{\mu d\over\sqrt{1-c_0^2}}F(\Phi,k).\eqno(3.28)$$
Here the quartic is given by
$$G(r)=(1-c_0^2)r^4-2c_0c_1r^3-(d^2+c_0d^2+c_1^2+2c_0c_2)r^2-(c_1d^2+2c_1c_2)r-c_2d^2
-c_2^2,\eqno(3.29)$$
the Jacobian $\mu$ and the parameter $k$ are the same as before (3.13) (3.15). The four
zeros of $G(r)$ are obtained by solving the two quadratic equations
$$1+u(r)=0,\quad r^2(1-u(r))-d^2=0.$$
This leads to
$$a_1={c_1\over 2(1-c_0)}+\sqrt{{d^2+c_2\over 1-c_0}+{c_1^2\over 4(1-c_0)^2}}$$
$$a_2=-{c_1\over 2(1+c_0)}+\sqrt{{c_1^2\over 4(1+c_0)^2}-{c_2\over 1+c_0}}$$
$$a_3=-{c_1\over 2(1+c_0)}-\sqrt{{c_1^2\over 4(1+c_0)^2}-{c_2\over 1+c_0}}\eqno(3.30)$$
$$a_4={c_1\over 2(1-c_0)}-\sqrt{{d^2+c_2\over 1-c_0}+{c_1^2\over 4(1-c_0)^2}}.$$
Then the exact lens equation is contained in the analogous equation to (3.19)
$$\pi-\tilde\beta={\mu d\over\sqrt{1-c_0^2}}[F(\Phi_{ds},k)+F(\Phi_d,k)].\eqno(3.31)$$
The angles $\Phi_d, \Phi_{ds}$ are given by the same formula (3.20) as before.
The appropriate expansion of the lens equation (3.31) depends on the particular values
$c_0, c_1, c_2$ in (3.27).

Regarding applications of our results one must replace the euclidean distances by
angular diameter distances as usual. Galaxies with joint lensing and dynamical data
can be found in the Sloan Lens ACS Survey (SLACS) and its follow-up project [8].
Unfortunately, until today only one system SDSSJ 2321-097 has been analyzed in detail.
This is an early-type elliptic galaxy which cannot be approximated by a spherically
symmetric metric. So we must extend our model-independent analysis the the elliptical
case or hope that the astronomers come up with a E0 lens galaxy.

\end{document}